# Influence of Bi addition on the property of Ag-Bi nano-composite coatings

Yuxin Wang[1, 2], See Leng Tay[2*], Xiaowei Zhou[1], Caizhen Yao[1, 2], Wei Gao[2], Guang Cheng[3*]

[1] School of Materials Science and Engineering, Jiangsu University of Science and Technology, Zhenjiang, 212003, Jiangsu, China

[2] Department of Chemical & Materials Engineering, the University of Auckland, PB 92019, Auckland 1142, New Zealand

[3] Physical and Computational Sciences Directorate, Pacific Northwest National Laboratory, P.O. Box 999, Richland, WA, 99352, USA

**Abstract**

Silver (Ag) coatings have been widely used in many industry areas due to their excellent conductivity. However, wider applications of Ag coatings have been hindered by their poor mechanical properties. In this research, to improve the mechanical performance, Ag-Bi nano-composite coatings were prepared by a novel ionic co-discharge method. A systematic study of the microstructure, mechanical properties, electrical conductivity and antibacterial behavior of the resulting coating was performed. The results indicated that after adding an appropriate amount of Bi containing solution into the Ag plating solution, Ag-Bi nano-particles were in-situ formed and distributed uniformly throughout the coating matrix, resulting in a significant improvement in the mechanical properties. The hardness of Ag-Bi coating was increased by 60% compared to that of the pure Ag coating. The corrosion resistance of Ag-Bi coatings was also enhanced. The outcome of this research may find a broader application in electronics, jewelry, aerospace and other industries.

---

[*] Corresponding authors.
Email address: seeleng.tay@cirrusmaterials.com (See Leng Tay) septem88@hotmail.com (Guang Cheng).

# 1. Introduction

Silver (Ag) or silver coatings have been widely used in diverse applications for their unique properties, such as shiny color, excellent electrical conductivity, high thermal conductivity and good antimicrobial properties [1-4]. For example, Ag coating/ film can be used as conductors for electrodes and ceramic capacitors in electronics, etc.[5] However, bulk Ag or Ag coatings are relatively soft, and the wear resistance of pure Ag is always much lower than that of nickel (Ni) or copper (Cu) [3, 6, 7]. Significant efforts have been made to improve the wear resistance (or enhance the hardness of the Ag coating), for example by introducing hard nanoparticles to generate Ag-based composites [8-10], creating new Ag-based alloys with special structures [1, 7, 11-15] and by reducing the grain size [11, 16-20]. Also, multiple deposition techniques, i.e., electrodeposition, physical vapor deposition (PVD), chemical vapor deposition (CVD) and magnetron sputtering, have been employed to prepare Ag coatings with improved properties [21]. Among those techniques, electrodeposition method has attracted plenty of interest due to its time efficiency, low cost, simple process and controllable micro/nanostructures [9, 16, 22].

An ideal Ag based coating should have good electrical conductivity and high wear resistance/hardness. As electrical conductivity and alloy strength are opposing properties, it is difficult to obtain an Ag coating with good electrical conductivity and high hardness simultaneously. Coating with a high proportion of alloying elements can significantly improve the hardness of Ag coatings but in most cases severely decrease electrical conductivity [21]. Ag nano-composite coatings have been extensively explored by introducing solid hard particles in the plating solutions. However, limited improvements were obtained as the second strengthening phases are tend to agglomerate during the plating process. We recently developed a new process, namely the ionic co-discharge method, which creates a composite coating without adding particles to the bath. Here the second phase material exists in the form of aqueous solution during the plating process. The precursor of nanoparticles exist as ions in the electrolyte and allows them to discharge onto the cathode, therefore avoiding the agglomeration of nano-particles effectively.

The solubility of bismuth in silver is relatively low (~1.5 at. %) according to it's phase diagram [22]. It has been shown that Bi addition is a promising method for improving the mechanical and physical performance of Ag coating [23]. In this research, Ag-Bi coatings with different quantities of Bi were prepared by our newly developed ionic co-discharge electrodeposition method. A systematic study was conducted in order to investigate the effects of different Bi addition on the performances of Ag-Bi nano-composite coatings.

## 2. Experimental
### 2.1 Sample preparation
The coatings were electroplated onto Ni-coated brass plates ($20 \times 25 \times 0.6$ mm$^3$). The purpose of the ~7 μm Ni layer is to prevent inter-diffusion between the brass substrate and Ag. Before electroplating, all specimens were pretreated in an alkaline solution at 80 °C for 10 seconds. A short Ag strike coating was applied for a period of 5 seconds, using stainless steel as the anode. The bath chemical composition and plating parameters are presented in Table 1. The Bi electrolyte was prepared as reported in our previous papers [23]. Fig. 1 depicts the schematic process of Ag-Bi nano-composite electroplating. With respect to the different Bi electrolyte quantities in the plating bath, the four coatings are designated as Ag, Ag+2.5 mL/L Bi, Ag+5.0 mL/L Bi and Ag+10.0 mL/L Bi in the remainder of this paper.

### 2.2 Sample characterizations
The crystal structures of coatings were measured by X-ray diffraction (XRD) with Cu Kα radiation (D2 Phaser Bruker, V=30 *KV*, I= 10 *mA*). Diffraction patterns were recorded in the 2 theta from 35º to 85º at the scanning step of 0.01º. The surface morphology and composition of coatings were analyzed using a field emission scanning electron microscope (FESEM) equipped with an energy dispersive spectroscope (EDS).

Cross-section images of the coatings are recorded after standard polishing process. Since the coating was expected to have a quite small grain microstructure, with grain sizes less than one micrometers, TEM images would be appropriate to provide further information about the coatings structure. The peeled off coatings were thinned using ion beam milling system, and the TEM images were taken by a high-resolution TEM (FEI TECNAI G2 F20, 200 kV).

The nanoindentation tests were conducted using a Hysitron TI950 Triboindenter in two directions (i.e., longitudinal and transverse directions) with a diamond Berkovich tip. The maximum indentation loading was 5 *mN* so that the maximum indentation depth was around 300 *nm* in both directions to avoid the substrate effect. The maximum load was held for 10s to avoid creep behavior in either indentation direction. Multiple indents were performed, and the corresponding load-depth curves were obtained. The surfaces perpendicular to the transverse direction were polished using a standard method so the influence of surface roughness could be ignored. The indentation on the as-deposited surface could be influenced by the surface roughness, so special care was taken to select indents and calculate the hardness following our earlier method [23, 24].

The electrical resistivity of coatings was measured by the four-point probe method. Before conducting the test, Ag and Ag-Bi coatings were peeled off from the substrate and were placed on a silicon substrate. Then the electric current was applied through the two outer probe points, and the potential was measured between two inner probe points with a Keithley 2602 Meter (Cleveland, OH, USA). The electrical resistivity of coatings were obtained by calculation using a standard method [25].

To evaluate the antibacterial behaviors of coatings, inhibition percentage tests were performed on coating samples. Escherichia coli (E. coli ATCC25922) which have been extensively studied were used. A colony E. coli was first cultured in 100 mL liquid Tryptic Soy Broth (TSB) medium overnight at 37 °C. Then 1 mL of E. coli containing the TSB medium was inoculated into 100 mL fresh TSB medium in sterile conic flasks.

Sterile conic flasks containing a 100 mL diluted E. coli solution and coatings samples were incubated at 37 °C for 18 hours. The control sample, without a coating sample, which contained only 100 mL E. coli solution, was also incubated. After incubation, the optical density of E. coli solutions was measured with a UV-Vis spectrophotometer (Agilent 8453) at a wavelength of 600 nm. All the antibacterial experiments were conducted in triplicate.

## 3. Results and Discussion
### 3.1 Phase structure of coatings

The XRD spectra of four coatings are shown in Figure 2. Five peaks are observed among the different coatings: Ag (111), Ag (200), Ag (220), Ag (311) and Ag (222). There is neither a Bi peak nor a metastable phase of $AgBi_2$ observed in the XRD spectra due to the low quantity Bi addition to the bath and consequent low amount of Bi dissolved in the Ag matrix. The influence of Bi addition on the patterns of the dominate peaks, Ag (111) and Ag (200), were shown in Figure 2(b) and (c), respectively. With the addition of Bi electrolyte, these two Ag peaks shift slightly to the lower diffraction angle: the greater the Bi electrolyte quantity, the greater the displacement shift. This would suggest that the addition of Bi generates new Ag-Bi solid solutions within the Ag matrix. An earlier study by Chithra et al. [13] also reported this same shift for Ag (111) when they prepared Ag-5.1% Bi alloy using mechanically milling.

### 3.2 Surface morphology and cross-section of coatings

Figure 3 shows the surface morphologies of Ag and Ag-Bi nano-composite coatings. All coatings show a typical nodular structure. The nodule shape with Ag-Bi coatings is more pronounced with the nodule size increasing as the Bi content increases.

The cross-sectional images of Ag and Ag-Bi nano-composite coatings are shown in Figure 4. Clear interfaces can be observed between Ni layer and Ag or Ag-Bi coatings. No cracks or abruptions were detected at the interface of the coatings, indicating a good adhesion between the Ni coated substrate and coatings. The three Ag-Bi coatings (~10 μm) were slightly thicker than the pure Ag (~8 μm) coating under the same electrodeposition conditions and operating parameters. All the Ag-Bi coatings have a similar thickness regardless of the Bi content. This comparison would indicate that alloying of Bi raised the deposition rate. Meanwhile, the weight percentage (wt. %) of Bi in the three Ag-Bi coatings were determined by EDS, as shown in Figure 5. A linear relationship was observed for the wt. % of Bi within the Ag-Bi coatings and the quantity of added Bi solution.

The TEM images of Ag and Ag-Bi nano-composite coatings are shown in Figure 6. The equiaxed grains were observed in the plane view TEM image of the pure Ag coatings and the grain size was ~300 nm as indicated in the highlighted white box in Figure 6 (a). The nano-sized twins with different thickness were also observed in the Ag coatings.

With the addition of Bi to the Ag matrix, smaller equiaxed grains are obtained as shown in Figure 6 (b), (c) and (d), indicating a grain refinement of coating matrix. Meanwhile, the number of nano-sized structures within grains were significantly reduced with the Bi addition. The pure Ag coatings carried fewer defects than the coating with Bi addition, consequently the TEM image shown in Figure 6 (a) is most simple and clean. Numerous internal defects were observed in TEM image of the Ag+5mL/L Bi coating and, as shown in Figure 6 (c), a large number of nanoparticles were observed in the center area of coating. In our previous study of an Ag+2.5mL/L Bi coating [24], high magnification TEM images demonstrated that nanoparticles with a size of ~10 nm are formed by the electrodepositing process. Hence, the Ag-Bi nano-composite coatings were successfully prepared in the current study.

### 3.3 Mechanical property of coatings

Nanoindentation is a small-scale in-situ testing and characterization method which has been widely used for coatings, thin films and bulk polycrystalline materials [26-32]. The load-depth (P-H) curves from the nanoindentation along the transverse direction (or on the cross-section view) the four coatings are presented in Figure 7. With respect to the influence of the possible microstructural anisotropy and texture of the coating layers [23, 24, 33, 34], the columnar or lamellar microstructures of the coating layers naturally give rise to the directionally dependent elastic-plastic properties. The anisotropic properties of the coatings or thin films could be characterization using either nanoindentation or resonance ultrasound spectroscopy [35-37]. In the current study, the nanoindentation tests were also conducted in the longitudinal direction (on the as-deposited surface). The hardness were calculated using the Oliver-Pharr method, and the results were shown in Figure 8.

With the same maximum indentation load, a higher indentation depth obtained on a coating indicates a softer alloy. Among the four coatings, the pure silver coating was the softest and the Ag+2.5 mL/L Bi is the stiffest. Despite the additional Bi content, the P-H curves of the two samples, Ag+5 mL/L Bi and Ag+10 mL/L Bi, were quite close to those of the Ag+2.5 mL/L Bi coating. The averaged hardness of the composite Ag coatings in the two directions was increased from ~1.65 GPa to ~2.67 GPa (Ag and Ag+2.5 mL/L Bi). This demonstrated that the reported ionic co-discharge method with a small volume of Bi electrolyte greatly improved the hardness of the Ag coating. Meanwhile, the large hardness difference observed between two directions in the Ag coating was greatly reduced after the Bi addition. The hardness difference in two direction among the three Ag-Bi nanocomposite coatings are quite small as shown in Figure 8. The volume loss during sliding wear is inversely proportional to the hardness of the coating according to Archard's law [38]. Consequently, in real world application as a contact marerial, at least 30% lower volume loss can be expected for the current Ag-Bi coatings in the compared to that of the pure Ag coatings.

As seen from the 3 Ag-Bi coatings, higher Bi addition might not be helpful to further improve the hardness of the coatings. As mentioned above, the Bi addition increased the deposition rate so the thickness of Ag-Bi coatings under identical conditions is greater than a pure Ag coating. The presence of Bi could also slightly affect the grain size of the Ag matrix. A comparison between Figure 6 (b), (c) and (d), show that a higher Bi addition increased the grain size of the Ag coatings. While there is almost no nano-twinned structure observed in the Ag+2.5 mL/L Bi coating, unlike the pure Ag coating, further addition of Bi electrolyte returns the nano-twinned structure to the nano-composite coating. In our previous study [23], the nano-twinned structure leads a lower yield strength and high hardening exponent in a pure Ag coating. Thus, the existence of nano-twinned structure results in a lower hardness for both the Ag+5 mL/L Bi and Ag+10 mL/L Bi. Then, the higher disparity between hardness values of the transverse and longitudinal directions is also observed with more Bi electrolyte. This correlation indicates that the addition of Bi electrolyte beyond 2.5 mL/L would influence the grain structures. Further study will be planned to analyze the effects of different Bi electrolyte volume on the grain growth and the grain morphologies.

**3.4 Electrical resistivity of coatings**

The electrical resistivity or the electrical conductivity is a crucial parameter for application of Ag in the electric industry. The electrical resistivity and conductivity of four coatings are listed in Table 2. Since the International Annealed Copper Standard (IACS) is a commonly used term to represent the conductive capacity of alloys, the electrical conductivity of the current coatings are also converted to the IACS in Table 2. A 100 % conductivity sample has the electrical resistivity as $1.7241\times10^{-8}$ $\Omega\cdot$m (or equal to the conductivity of $5.80\times10^{7}$ S/m) at room temperature [39].

The pure Ag coating possesses the highest electrical conductivity among the four coatings. The addition of Bi to Ag matrix did slightly increase the electrical resistivity as expected. A general relationship has been summarized in our previous study [23] that the higher volume of alloying elements, the higher the electrical resistivity. This relationship is very consistent in the binary alloy systems of Ag when the atomic percent alloying elements is no more than 20%. As the compassion between Ag and Ag+2.5 mL/L Bi, the ~6% increase in the electrical resistivity is quite small compared to the 60% increase in the hardness. The low electrical resistivity (at least 87% IACS) and the high hardness of Ag-Bi coating demonstrated that the current ionic co-discharge method successfully prepares a much stiffer coating with minimal reduction in electrical conductivity. The Ag-Bi nano-composite coating is a good candidate for electrical contacts.

**3.5 Antimicrobial property of coatings**

The antimicrobial behaviors of four coatings were investigated using *E. coli* ATCC 25922. The inhibition percentage of the coatings was calculated by using the equation

$$I\% = \frac{(con_{18} - con_0) - (samp_{18} - samp_0)}{(con_{18} - con_0)} \times 100$$

where I % is the percentage inhibition of growth, $con_{18}$ and $con_0$ are the optical density at 600 nm of *E. coli* in TSB as the control of the organism at 18 h and 0 h, respectively. $Samp_{18}$ and $samp_0$ are the optical density at 600 nm of *E. coli* in TSB with the presence of samples at 18h and 0 h, respectively.

Then, the inhibition percentage of four coatings after 18 hours was calculated and listed in Table 3. The antimicrobial property of the substrate (Ni coated brass) was also tested as a comparison. The inhibition percentage of *E. coli* growth for pure Ag coating is 76.6 %, which is much higher than the 31.2% measured from the substrate.

The inhibition percentage of Ag-2.5 mL/L Bi coating is 76.2 % which is quite close to that of the pure Ag. Further addition of Bi led to a slight decrease of the inhibition for bacteria growth. However, strong inhibition capability (>70 % inhibition) was observed among different Ag-Bi nano-composite coatings when compared to the substrate.

Silver contains a low toxicity and can be used as effective antibacterial agents in medical fields [40, 41]. Several mechanisms have been proposed for the antibacterial mechanism of Ag ions, i.e., protein dysfunction; production of reactive oxygen species and depletion of antioxidants; impaired membrane function, and interference with the nutrient assimilation [42, 43]. Protein dysfunction is the main antibacterial mechanism. In the case of Ag coating, Ag ions bind to electron donor groups such as sulfhydryl, amino, and carboxyl of proteins, leading to their denaturation and consequently to the inactivation of several vital functions of the bacteria [44].

Multiple factors, i.e., the concentration of the ion release [45], surface morphology/roughness [46], grain size [47], etc., will determine the antibacterial behaviors of a coating. As the SEM images shown in Figure 4, the surface roughness would barely influence with Bi addition. Meanwhile, the smaller Bi concentration would affect the concentration of the ion release. Hence, the grain size and different grain cluster shown in Figure 3 would be the reason lead to different antibacterial behaviors.

**4. Conclusions**

In the current study, we utilized a novel ionic co-discharge method to prepare high-performance Ag-Bi nanocomposite coatings. A small addition of Bi electrolyte greatly raises the hardness of the coatings while maintaining the high electrical conductivity and high antibacterial capability of pure Ag. Different characterization methods, i.e., XRD, SEM, and TEM demonstrated the nano-composite structures of Ag-Bi coatings. The outcome of this research will have a broader application in electronics, jewelry, aerospace, and other industries.

**Acknowledgments**

The University of Auckland performed all the experimental work reported in this work, and it was supported by National Natural Science Foundation of China (51601073), Jiangsu Distinguished Professor Project (1064901601) and Auckland UniServices Project. Pacific Northwest National Laboratory (PNNL) is operated by Battelle Memorial Institute for the US Department of Energy (DOE) under Contract No. DE-AC06–76RL01830. The authors would like to thank the technical stuff in the Department of Chemical and Materials Engineering and the Research Centre of Surface and Materials Science for various assistance. We also want to express our gratitude to Mr. Glen Slater, Mr. Chris Goode and technical staff in Rigg Electroplating Ltd.

**Tables**

**Table 1** Solution composition and processing parameters of electrodeposited Ag and Ag-Bi coatings.

| Bath composition & plating parameters | Quantity |
|---|---|
| Silver metal | 30 g/L |
| KCN free | 120 g/L |
| KOH | 10 g/L |
| $Bi(NO_3)_3 \cdot 5H_2O$ | 97 g/L |
| Tartaric acid | 30 g/L |
| KOH | 140 g/L |
| Temperature | Room temperature (25 °C) |
| Current density | 10 mA/cm$^2$ |
| Agitation speed | 200 rpm |
| Plating time | 30 min |
| Bi electrolyte | 2.5 mL/L, 5.0 mL/L & 10.0 mL/L |

**Table 2** The electrical resistivity and electrical conductivity of four coatings

| Coating | Resistivity ($\times 10^{-8}$ $\Omega \cdot$m) | Conductivity ($\times 10^7$ S/m) | Conductivity % IACS |
|---|---|---|---|
| Ag | 1.78±0.02 | 5.63±0.07 | 97.1±1.2 |
| Ag+2.5 mL/L Bi | 1.88±0.02 | 5.31±0.06 | 91.5±1.0 |
| Ag+5 mL/L Bi | 1.92±0.01 | 5.22±0.03 | 89.9±0.4 |
| Ag+10 mL/L Bi | 1.99±0.02 | 5.04±0.04 | 86.8±0.7 |

**Table 3** The percentage of inhibition for four coatings

| Coating | Percentage of Inhibition (%) |
|---|---|
| Substrate (Ni coated brass) | 31.2 |
| Ag | 76.6 |
| Ag+2.5 mL/L Bi | 76.2 |
| Ag+5 mL/L Bi | 72.8 |
| Ag+10 mL/L Bi | 71.6 |

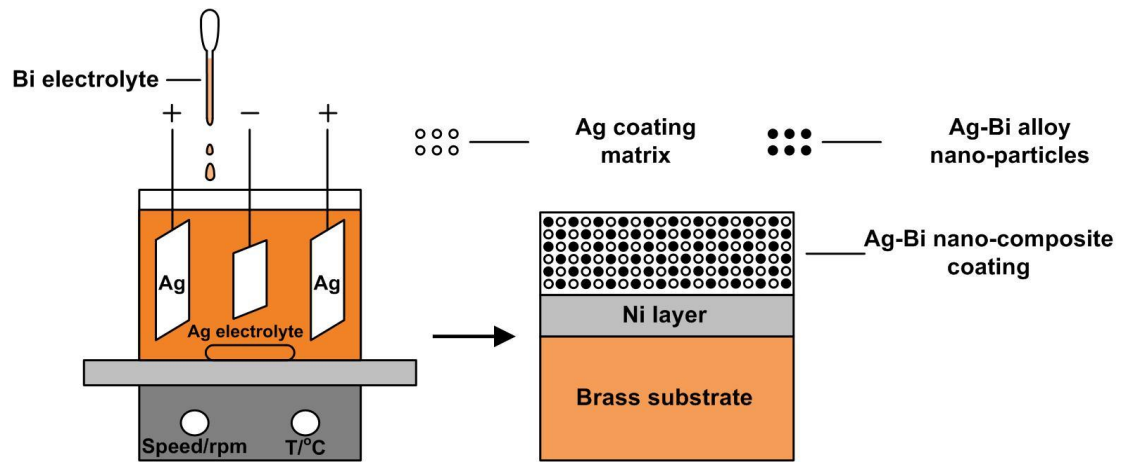

Figure 1: The schematic process of Ag-Bi nano-composite electroplating

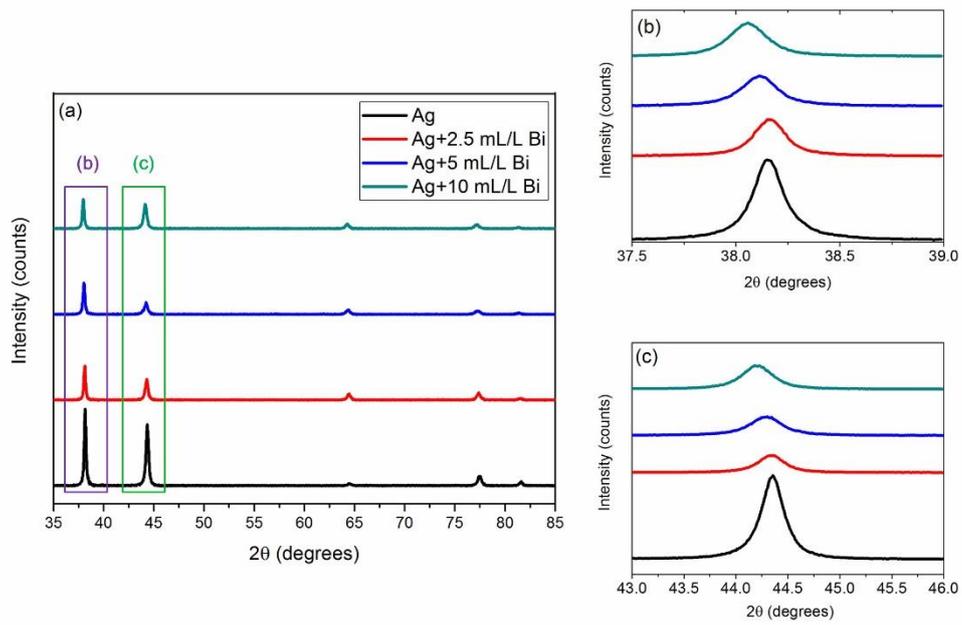

Figure 2: XRD patterns of (a) four coatings with different amount of Bi, (b) magnified peak of Ag (111) and (c) Ag (200)

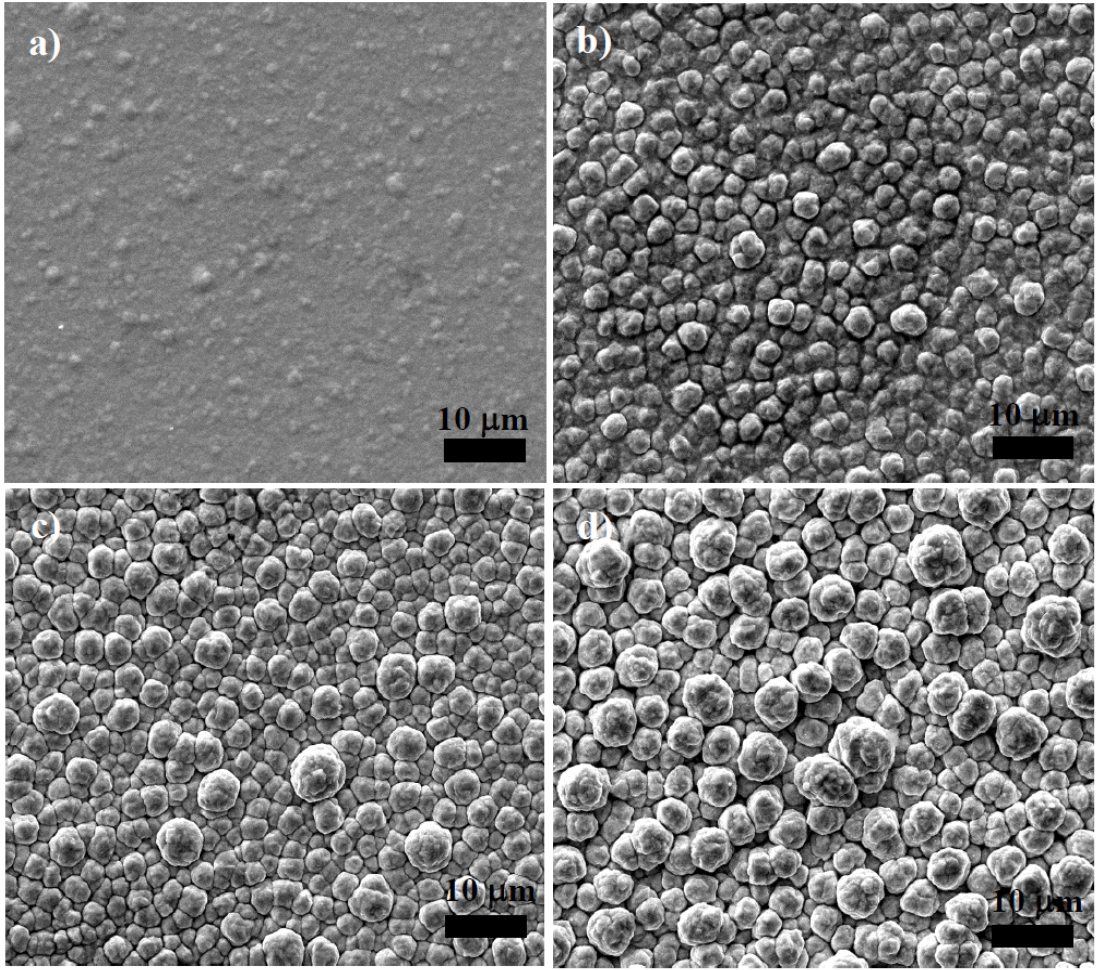

Figure 1: The morphologies of as-deposited surfaces under SEM: (a) Ag, (b) Ag+2.5 mL/L Bi, (c) Ag+5 mL/L Bi, and (d) Ag+10 mL/L Bi

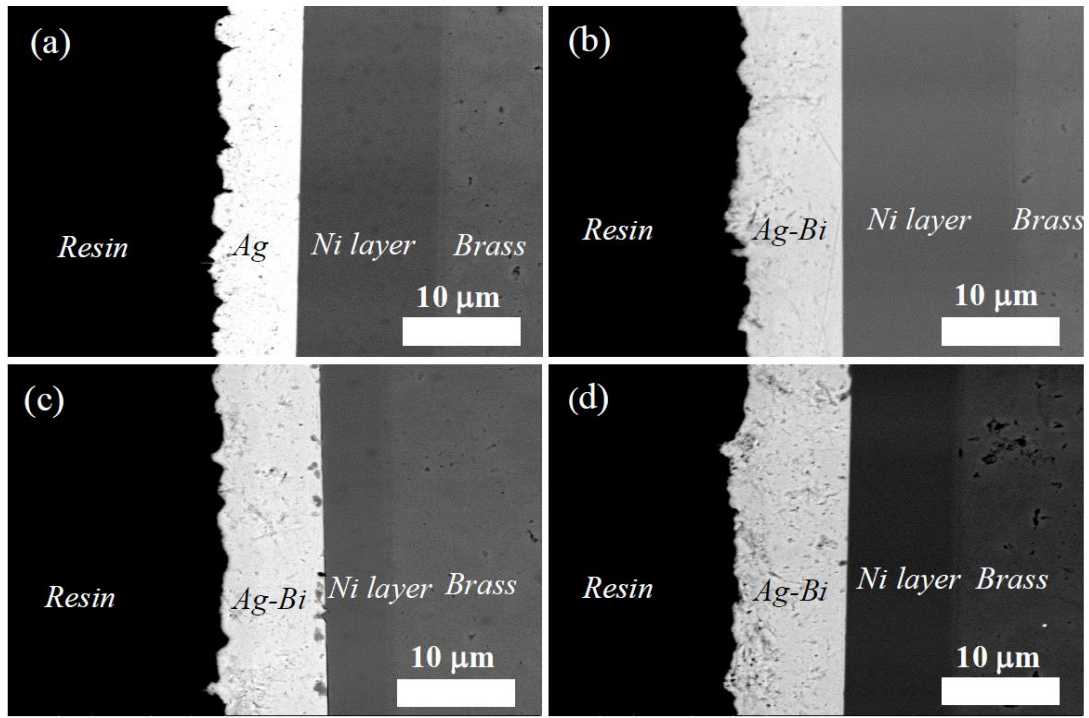

Figure 2: The cross-section images for four samples under SEM: (a) Ag, (b) Ag+2.5 mL/L Bi, (c) Ag+5 mL/L Bi, and (d) Ag+10 mL/L Bi

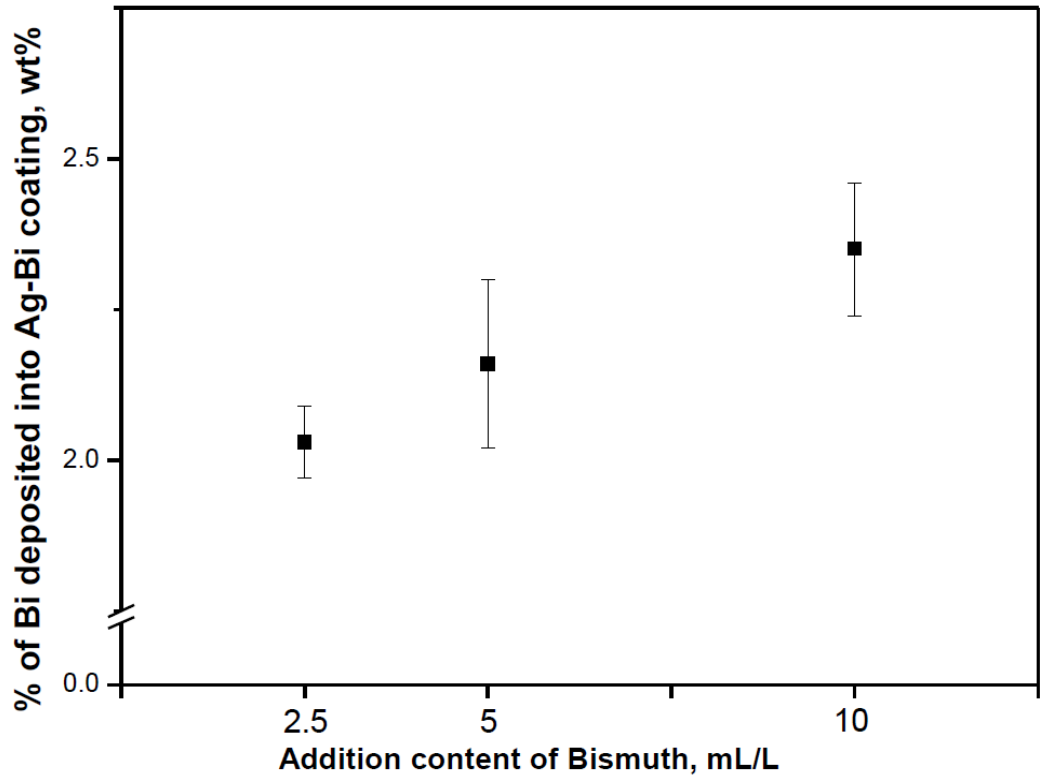

Figure 5: The correlation between wt. % of Bi in coatings and the addition content of Bi

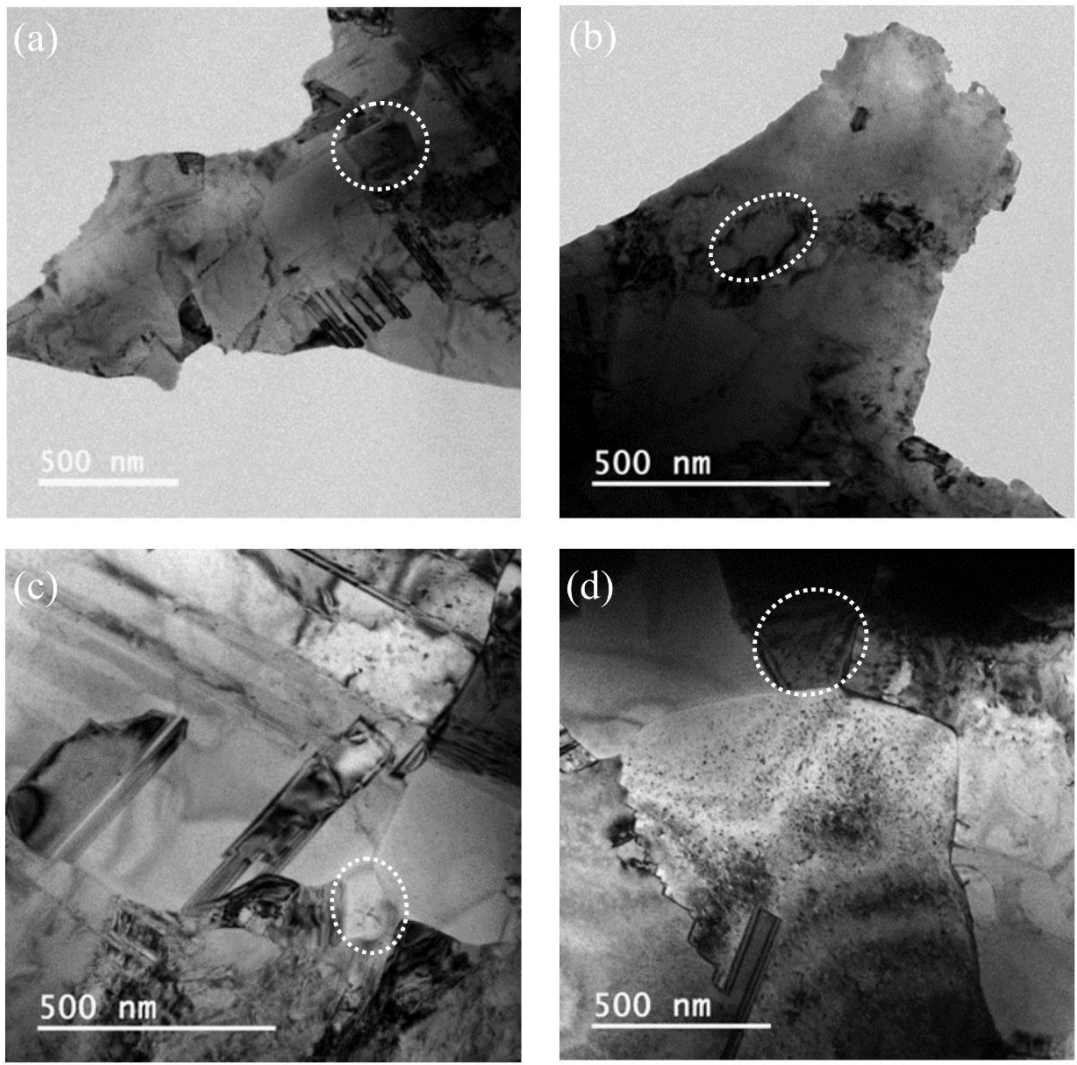

Figure 6: TEM images of the nanocrystalline coatings from the plane view with individual grains highlighted in the white dotted box: (a) Ag, (b) Ag+2.5 mL/L Bi, (c) Ag+5 mL/L Bi, and (d) Ag+10 mL/L Bi

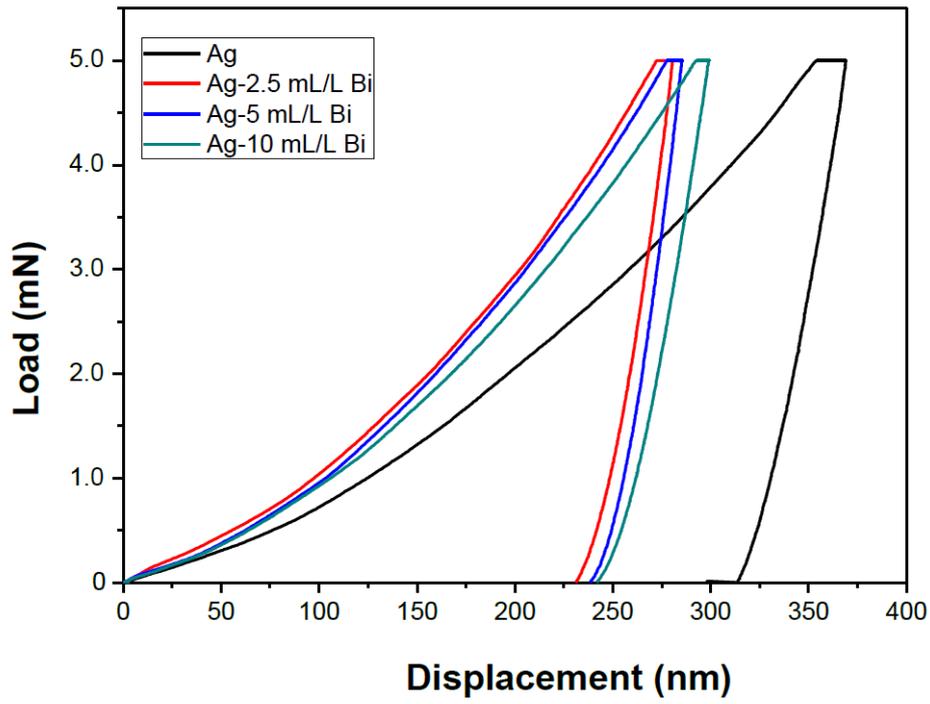

Figure 7: The load-depth curves of four coatings on the cross-section surface with the same indentation load

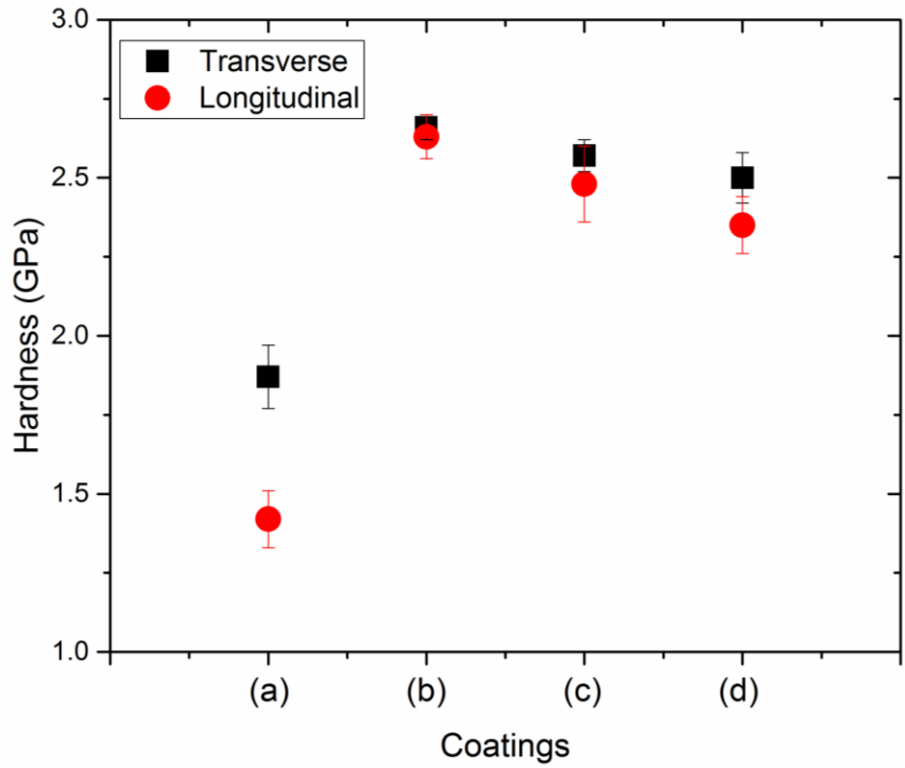

Figure 8: The hardness values of four coatings on transverse direction and longitudinal directions: (a) Ag, (b) Ag+2.5 mL/L Bi, (c) Ag+5 mL/L Bi, and (d) Ag+10 mL/L Bi